# IN MEMORY OF JULIAN SCHWINGER

Mario Rabinowitz
Armor Research
715 Lakemead Way, Redwood City, CA 94062-3922
Mario715@earthlink.net

Although he was the recipient of a Nobel Prize and despite the greatness of his accomplishments, Julian Schwinger is almost an unsung hero of our age . He is relatively unknown to the general population, even though in the physics community he was a renowned theoretician and teacher of physics. He shared the 1965 Nobel Prize in physics with Richard P. Feynman and Shin'ichiro Tomonaga for their development of Quantum Electrodynamics (QED). Of these three extraordinary physicists, and even of all the physicists that worked on QED, his work was the most rigorous and mathematically exacting.

Schwinger received his Bachelor of Arts degree from Columbia University in 1936 at the age of 18, where his phenomenal skills in physics were quickly recognized by his professors. He went on to get his PhD in physics in record short time at Columbia at the age of 21. His genius was clearly recognized by the elite members of the physics community. Even before he received his doctorate, he co-authored a paper with Edward Teller; and such other notable physicists as Enrico Fermi and Hans Bethe also collaborated with him.

Because of his early achievement and recognition, it may seem that it was all smooth sailing through college for this prodigy, but it was not. He came close to failing undergraduate chemistry at Columbia, and it is rumored that he almost didn't get his doctorate since he didn't want to be encumbered by having to pass the tough reading exam in German. Because of his exceptional ability in physics, this requirement was waived. Actually, if it weren't for his mentor and patron, Nobel Laureate I. I. Rabi, he might not even have earned his bachelor's degree.

In 1934, at the age of 16, Schwinger entered the City College of New York. At City College, he greatly impressed Professor Mark Zemansky, by correctly working every problem in Zemansky's *Heat and Thermodynamics* book. Apocryphal stories abound that he rarely attended classes, and never finished a lab experiment -- after all, he didn't set out to become an experimentalist. What is certain is that with the exception of a few professors who recognized his genius and supported him, Schwinger was in administrative trouble. Rabi came to the rescue, and got him transferred to Columbia University. Before getting his bachelor's degree, he essentially finished the research that was to become his doctoral thesis.

Mainly self-educated, he wrote his first paper in 1934 which was published by the time he was 17. It was "On the Interaction of Several Electrons," remotely related in content to his Nobel Prize work on QED which crystallized 14 years later. His first really important paper was published in *Physical Review* in 1937, just before his nineteenth birthday, on the scattering of neutrons by magnetic materials. I think he was the first to suggest scattering of neutrons from ortho- and para-hydrogen to determine the neutron spin.

His first teaching position was at Purdue University from 1941 to 1943, where he went from instructor to assistant professor. He is still fondly remembered there. Purdue recently gave him an honorary doctorate. The old-timers still recall how he kept his automobile clean and shiny, polishing it almost daily. He may not have been able to afford a Cadillac then, but a shiny Cadillac and immaculate attire became his hallmark.

From Purdue, Schwinger went on to the Radiation Lab at MIT. Here he published some of the finest work on synchrotron radiation. His work there on radar from 1943 to 1945 was important to the US in winning the Second World War. By this time his work was well known to an even larger segment of the physics community, and he accepted a position as Associate Professor at Harvard University in 1945. He was promoted to full professor two years later at age 29.

It is said that his class lectures verged on perfection and were delivered without ever looking at his notes. His lectures on nuclear physics were attended by students and faculty alike from both Harvard and MIT. Herbert Goldstein, professor of physics at Harvard and author of one of the foremost texts on classical mechanics, acknowledged his debt to Schwinger. Schwinger joined the faculty of the UCLA physics department in 1972 as University Professor, and maintained his affiliation there beyond his retirement. To Schwinger's great credit, he nurtured over 70 PhD students among whom were three Nobel Laureates: Ben Mottelson (physics, 1975), Sheldon Glashow (physics, 1979), and Walter Gilbert (chemistry, 1980). To my knowledge, this is the largest number of doctoral students, as well as the largest number of Nobel Prize winners, that any one professor has ever fostered. He not only did great science himself, but also inspired his students.

Like James Clerk Maxwell, Schwinger was unusually honest in his publications, even pointing out mistakes he had previously made. He freely gave seminal ideas to others which led to highly significant results and even the Nobel Prize without asking for credit. He suggested to Sheldon Glashow that the weak and electromagnetic forces might be unified into the electroweak force. He suggested to Timothy Boyer that the Casimir force might explain what holds an electron together. His most recent interests were cold fusion and sonoluminescence.

Schwinger was a pioneering theorist in cold fusion. He felt that the bias of the physics community against cold fusion was based on inferences from hot fusion that are not valid in this new regime. He argued that the defense of cold fusion can be simply stated, "The circumstances of cold fusion are not those of hot fusion." He first submitted the results of his theoretical analysis of cold fusion to journals of the American Physical Society. He received such harsh treatment in the denial of publication of this work that as a symbolic gesture, he resigned his membership from the American Physical Society. This was no small step for someone who had been a leading member for over 50 years. In doing so he said, "The pressure for conformity is enormous. I have experienced it in editors' rejection of submitted papers, based on venomous criticism of anonymous referees. The replacement of impartial reviewing by censorship will be the death of science."

Undaunted, Schwinger did publish papers on cold fusion in other forums. He was the sole author of the following 8 papers on this subject:
1. "Nuclear Energy in an Atomic Lattice," *Proceedings of the First Annual Conference on Cold Fusion* (Salt Lake City) pp. 130 - 136, March 28 -31 (1990).
2. "Nuclear Energy in an Atomic Lattice.1," *Zeitschrift fur Physik D* **15**, pp. 221-225 (1990).
3. "Cold Fusion: A Hypothesis," *Zeitschrift fur Naturforschung* **45a**, p. 756 (1990).
4. "Cold Fusion: Does It Have a Future?" in *Evolutional Trends of Physical Sciences.* Germany: Springer Verlag 1991. (From a talk delivered in Tokyo, 1990)
5. "Phonon Representations," *Proceedings of the National Academy of Sciences* **87**, pp. 6983 - 6984 (1990).

6. "Phonon Dynamics," *Proceedings of the National Academy of Sciences* **87**, pp. 8370 - 8372 (1990).
7. "Nuclear Energy in an Atomic Lattice - Causal Order," *Progress in Theoretical Physics* **85**, pp. 711 - 712 (1991).
8. "Cold Fusion Theory: A Brief History of Mine," *Fourth International Conference on Cold Fusion* (Maui, 1993) *ICCF-4 Transactions of Fusion Technology* pp. ? - ? (1994).

He also presented a colloquium at MIT and the University of Pennsylvania entitled, "A Progress Report: Energy Transfer in Cold Fusion and Sonoluminescence."

When I first read Schwinger's papers on cold fusion, I thought that I may have spotted some errors in them. So I wrote him a letter on September 24, 1990, respectfully raising a couple of questions. I did not hear from him directly, but did speak with his devoted wife, Clarise. Because of my great admiration for Professor Schwinger, in co-authoring two reviews of cold fusion (*International Journal of Theoretical Physics* **33,** pp. 617 - 670, 1994; and in ***Trans. Fusion Technol.* 26**, 3 (1994)) I felt badly about writing a critical review of his work. I wrote, "We hope that Schwinger will address the issues raised and clarify the situation." In hopes that he would resolve the questions that we raised, I again wrote to him in early 1994 sending him a copy of our paper. Perhaps because he was ill, he neither responded privately nor publicly. It would not have troubled me if he had proven me wrong.

I was profoundly saddened when I learned that he passed away on July 16, 1994. He was one of my heroes. Even though I did not know him personally, I feel a personal loss as well as a loss to the world of one of the greatest minds of our time.